\documentclass[3p,times,twocolumn,preprint]{elsarticle-mod}

\usepackage{ecrc-mod}

\firstpage{1}

\runauth{{Henryk Czy\.z} and {Sergiy Ivashyn}}

\usepackage{amssymb}

\usepackage{hyperref} 
\usepackage{amsmath,latexsym}
\usepackage{graphicx}

\begin{document}

\begin{frontmatter}




\vspace{-30pt}

\title{EKHARA --- new gamma gamma generator}


\author[label1]{Henryk Czy\.z}
\author[label2]{Sergiy Ivashyn~\corref{cor1}}
\address[label1]{A.~Che{\l}kowski Institute of Physics, University of Silesia, Katowice PL-40007, Poland}
\address[label2]{
ITP, NSC ``Kharkov Institute of Physics and Technology'', Kharkov UA-61108, Ukraine
}
\cortext[cor1]{Corresponding author}
\ead{s.ivashyn$\otimes$gmail.com}

\begin{abstract}
The Monte Carlo generator {\tt EKHARA} allows to simulate the process 
$e^+e^- \to e^+e^-\pi^0$ at energies of meson factories. 
We review the motivation, describe the current status of the generator, 
and discuss features, which will appear in the forthcoming releases: 
the two-photon production of $\eta$ and $\eta'$ mesons.
\end{abstract}


\end{frontmatter}


\section{Introduction}
\label{sec:intro}

\begin{figure*} \begin{center}
 \resizebox{0.35\textwidth}{!}{%
 \includegraphics{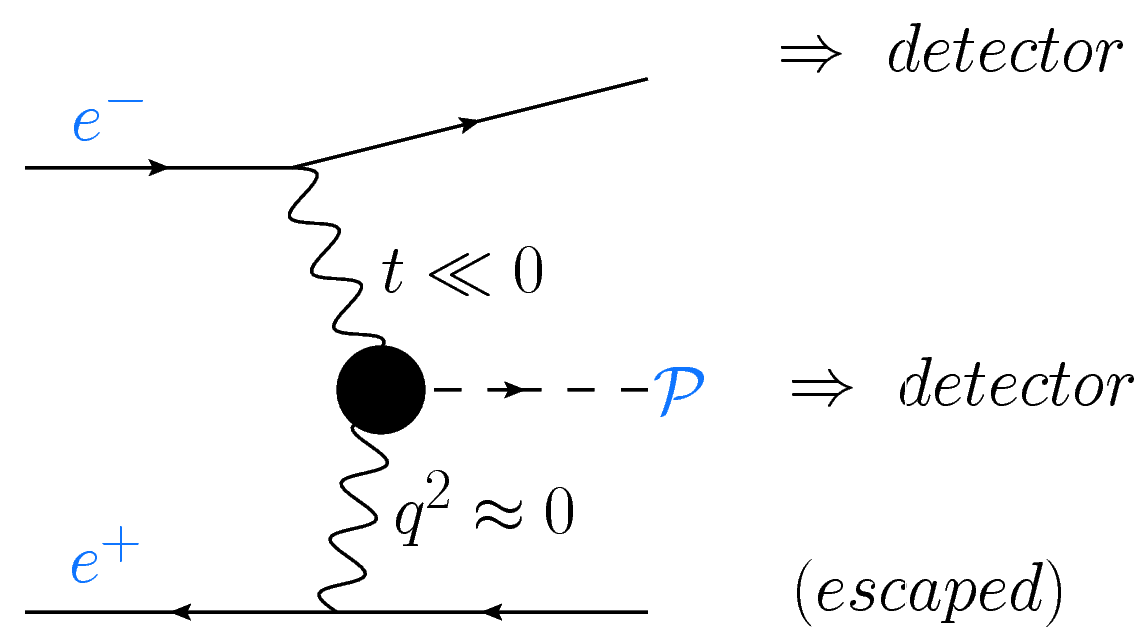}
  } 
\hfill  
 \resizebox{0.45\textwidth}{!}{%
   \includegraphics{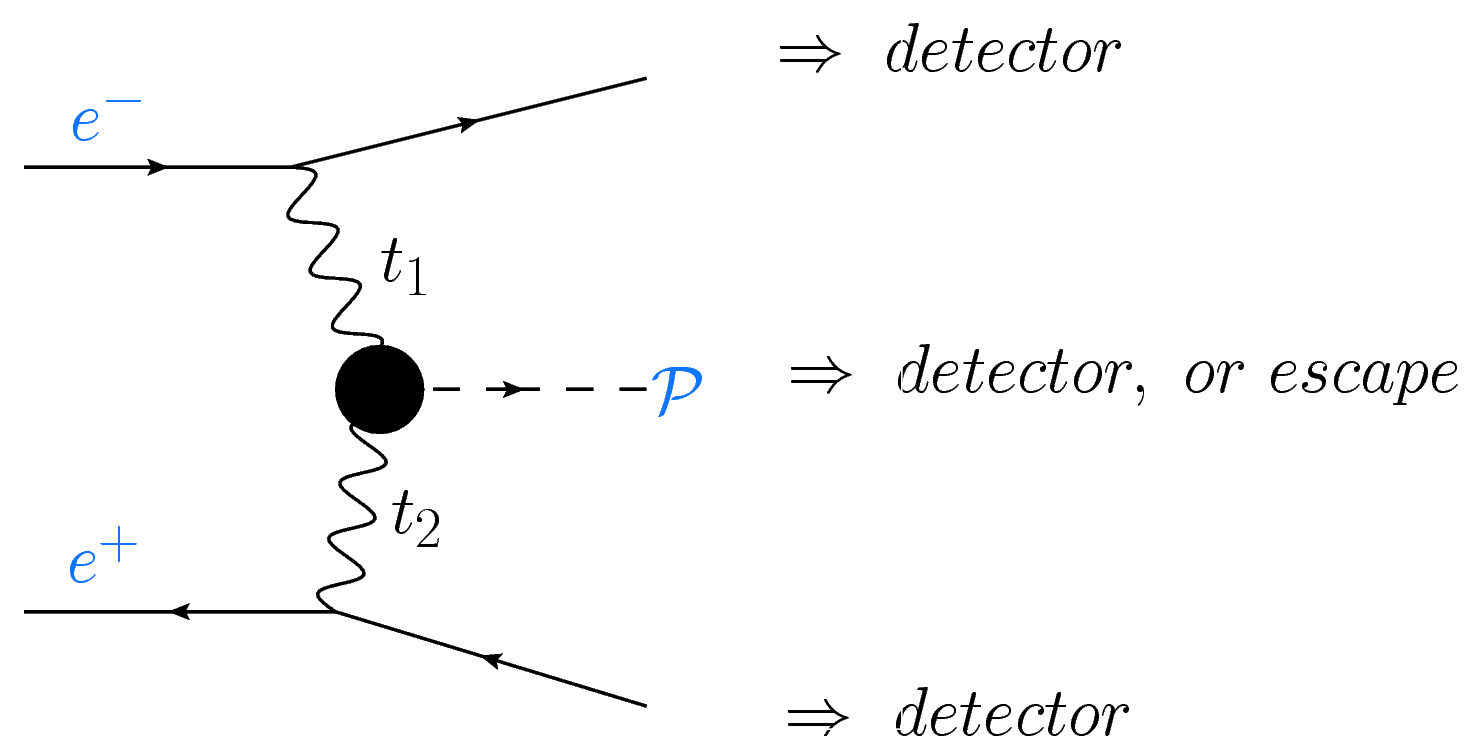}
 } 
 \end{center}
 \caption{ 
  The concept of single-tag (left) and
  double-tag (right) $e^+e^- \to e^+e^-\mathcal{P}$ experiment. 
 }
 \label{fig:single-double-tag}
 \end{figure*}

The two-photon transition form factors 
of the light pseudoscalar mesons ($\mathcal{P} = \pi^0$, $\eta$, $\eta^\prime$) 
are being intensively discussed during the last years.
Recently the BaBar experiment has provided us with new information in the region
where the virtuality of the first photon is high ($Q^2$) and that of
the second one is small (and both photons are spacelike)~\cite{Aubert:2009mc,:2011hk}.
The process, which gives an access to the given kinematic region, is 
$e^+e^- \to e^+e^-\mathcal{P}$ in the ``single-tag'' mode, 
see Fig.~\ref{fig:single-double-tag}~(left).
Within the range of the previous experiments (CELLO~\cite{Behrend:1990sr}, CLEO~\cite{Gronberg:1997fj}) there was a fair agreement with the theory.
The results of BaBar on the pion transition form factor~\cite{Aubert:2009mc}
demonstrated an unexpected deviation from the established theoretical predictions
and has triggered the reconsideration of the theoretical description of 
these form factors~\cite{Dorokhov:2009jd,Kroll:2010bf,Brodsky:2011xx,Brodsky:2011yv,Klopot:2011qq,Balakireva:2011wp,Bakulev:2011rp,Stefanis:2011kp,Dorokhov:2011dp,Bystritskiy:2009bk}.
Over the last year several theoretical advances appeared, which
were able (or at least claimed to be able) to reconcile 
theory and experiment (e.g.,~\cite{Kroll:2010bf,Noguera:2011fv})
for the $\pi^0$, $\eta$  and  $\eta'$ simultaneously.
Nevertheless, there are also indications
that the data of Ref.~\cite{Aubert:2009mc} on $\pi^0$ 
cannot be easily accommodated by
theory (see~\cite{Brodsky:2011xx,Balakireva:2011wp,Bakulev:2011rp,Stefanis:2011kp,Bakulev:2011iy,Lucha:2011if}).
Therefore, there are still open questions and 
further research is foreseen both on the theory and the experiment side.

The new experiments, e.g., KLOE-2~\cite{AmelinoCamelia:2010me,Babusci:2011bg},
and BES-III~\cite{Asner:2008nq} will soon start measuring the two-photon
form factors of the light pseudoscalars and related observables.
In case of a double-tag experiment, see Fig.~\ref{fig:single-double-tag}~(right), 
the form factor as a function of both photon virtualities is 
potentially of interest. 
Currently, the double-tag experiments are the one performed
with the KEDR detector at Novosibirsk~\cite{Zhilich:phipsi:talk}
and the one performed by KLOE-2 Collaboration at Frascati~\cite{AmelinoCamelia:2010me,Babusci:2011bg}.
Unfortunately, in the above experiments
the shape of the form factor as a function of both photon virtualities
is not accessible.
%
%
Notice that the form factor unknown dependence 
on the second photon virtuality has possible impact on the precision 
of the experimental extraction of the quoted values of the form factor
even in case of a single-tag experiment,
as has been stressed in~\cite{Lichard:2010ap}
and studied in~\cite{Czyz:2011:preparation}.
In any case, a realistic simulation has to account for
both photon virtualities.

The transition form factor description is not only a milestone in
our understanding of the $U(1)$ axial anomaly~\cite{Klopot:2011qq}, 
general principles of the strong interaction 
dynamics~\cite{Bakulev:2011rp} and $SU(3)$ flavor symmetry breaking~\cite{Feldmann:1997vc},
but it is also very important in the context of the Standard Model tests
at low energy~\cite{Dorokhov:2009jd}, 
mainly in the problem of the anomalous magnetic moment of 
the muon $a_\mu$. 
The transition form factors (though in a different kinematic region) 
are required for the calculation of the hadronic light-by-light (hLbyL) scattering
part of the $a_\mu$ usually denoted
as $a_\mu^{\mathrm{LbyL}}$~\cite{Melnikov:2003xd,Dorokhov:2008pw,Prades:2009tw,Nyffeler:2009tw,Jegerlehner:2009ry,Dorokhov:2011zf}.
In view of the new $g-2$ experiments to be carried out 
at Fermilab~\cite{Carey:2009zzb} and JPARC~\cite{JParc:amm:2010}, 
it is crucial to improve the accuracy of $a_\mu^{\mathrm{LbyL}}$~\footnote{
This subject has been a topic of the 
``INT Workshop on the Hadronic Light-by-Light Contribution to the Muon Anomaly''
(Seattle, February 28 -- March 4, 2011) 
\href{http://www.int.washington.edu/PROGRAMS/11-47w/}{http://www.int.washington.edu/PROGRAMS/11-47w/}}.
In context of hLbyL the improvement of the models for the two-photon form factors
is important, which requires the high precision of the form factor 
measurements~\cite{Babusci:2011bg}.
In turn, precision measurements require precision simulation.

Summing up, the state of the art is that both theory and experiment need a
computer program for the process $e^+e^- \to e^+e^-\mathcal{P}$
-- the Monte Carlo generator, which would allow to simulate the events 
under the conditions similar to the experimental ones.
Our objective is to develop and to maintain a free and public
tool for $e^+e^- \to e^+e^-\mathcal{P}$ simulation ($\pi^0$, $\eta$, $\eta^\prime$)
to be used by experimentalists and theorists.
The guiding principles are the following:
\begin{itemize}
\item to account for light meson phenomenology in detail
      via precise modeling of the form factors,
      in agreement with available data;
\item full description of both virtual photons;
\item high generator efficiency: 
      accounting for the peaking behaviour of the cross section.
\end{itemize}
This approach is being realized in the Monte Carlo generator
{\tt EKHARA}~\cite{Czyz:2010sp,Czyz:2006dm}~\footnote{
{\tt EKHARA} web page: 
\href{http://prac.us.edu.pl/~ekhara}{http://prac.us.edu.pl/\~\;\!ekhara}}.
In the following Section we review the current status of this program.

\section{{\tt EKHARA} Monte Carlo generator}

The Monte Carlo generator {\tt EKHARA} is a computer program 
written in {\tt FORTRAN} using the quadruple precision numbers.
The principal description of the program is given in 
the dedicated paper~\cite{Czyz:2010sp}.
The generator can be used a stand-alone code or be interfaced
to another (e.g., an experiment's) simulation program.

The first version of {\tt EKHARA}~\cite{Czyz:2005ab,Czyz:2006dm}
allows to simulate the reaction $e^+ e^- \to \pi^+ \pi^- e^+ e^-$ 
with the focus on a specific event selection, 
namely that of the radiative return method for 
cross section $\sigma(e^+ e^- \to \pi^+ \pi^-)$ measurement.
It was used for the background~\cite{Czyz:2003gb} simulation 
by KLOE Collaboration~\cite{Aloisio:2004bu,:2008en}.
This channel can be important for the radiative return program of 
the BES-III experiment~\cite{Actis:2010gg,Asner:2008nq}.
The $e^+ e^- \to \pi^+ \pi^- e^+ e^-$ mode in {\tt EKHARA}
has the following features, which a user should be aware of:
it generates weighted events and the applied phenomenology of 
the $\gamma^*\gamma^* \to \pi^+ \pi^-$ subprocess {\it is not}
tuned to reflect the line shapes known from the dedicated
$\gamma^*\gamma^* \to \pi^+ \pi^-$ experiments.
An improvement of the above issue is on the agenda.

The second version of {\tt EKHARA}~\cite{Czyz:2010sp}
features a new channel, $e^+e^- \to e^+e^-\mathcal{P}$.
%
The implementation of this channel 
and the applied phenomenology {\it is} oriented
on the $\gamma^*\gamma^*$ physics.
The phase space construction 
and the matrix element calculation is based on 
the exact formulae and the exact kinematics 
(in contrast to the Equivalent Photon Approximation,
which was used in many generators in this field).

\begin{figure} \begin{center}
 \resizebox{0.45\textwidth}{!}{%
 \includegraphics[width=0.45\textwidth]{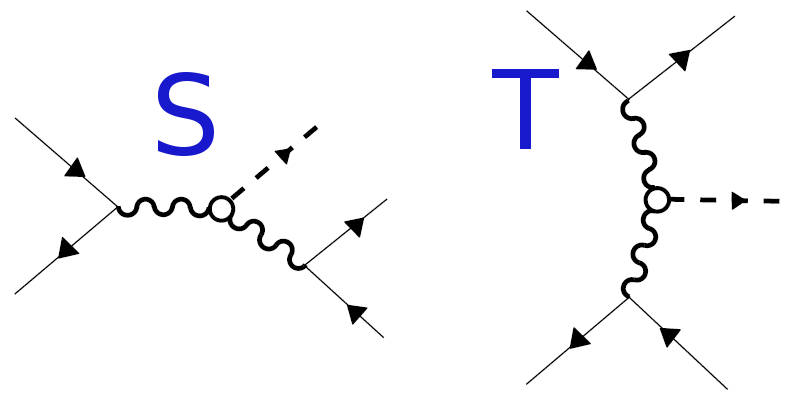} 
 } 
 \end{center}
 \caption{ 
 The $s$- and $t$-channel amplitudes in $e^+e^- \to e^+e^-\mathcal{P}$.
 }
 \label{fig:s-t}
 \end{figure}

It is important to stress that the (user-provided) 
form factors in {\tt EKHARA} have to be meaningful both in the time-like
and space-like regions of the photon, because
both $s$- and $t$-channel amplitudes and 
their interference can be switched on in the simulation,
see Fig.~\ref{fig:s-t}.
In general, the provided form factors have to be functions 
of {\it two} photon virtualities.
The above criteria considerably reduces the user's choice,
because the majority of the published formulae for the
form factors within different theoretical approaches 
hold only for the case with one photon being real and the other
--- space-like and virtual.
Strictly speaking, from the point of view of the numerical simulation,
such a case is very specific and is 
{\it never} realized for the $t$-channel amplitude
because the leptons are not massless.

In order to support a user, we have recently~\cite{Czyz:2011:preparation} supplied {\tt EKHARA} with 
the transition form factors for $\pi^0$, $\eta$ and $\eta^\prime$ mesons.
Inclusion of $\eta$ and $\eta^\prime$ meson production in {\tt EKHARA}
is a new feature with respect to~\cite{Czyz:2010sp}.
The form factor modeling~\cite{Czyz:2011:preparation} was performed in the framework 
of the effective field theory, namely the chiral theory with 
resonances~\cite{Ecker:1989yg,Ecker:1988te,Prades:1993ys} was used.
The quality of the model is good: by fitting only one
free parameter to the data for $\pi^0$, $\eta$ and $\eta^\prime$ form factors
(simultaneously) we achieved $\chi^2/d.o.f. \approx 1.2$.
The existing data from single-tag experiments in space-like region
(CELLO~\cite{Behrend:1990sr}, CLEO~\cite{Gronberg:1997fj} 
and BaBar~\cite{Aubert:2009mc,:2011hk}) were used not only
as a reference for model improvement, but also  
for the generator verification.
The verification is done by comparison of the simulated
cross sections $\frac{d\sigma}{d Q^2}$ 
for $e^+e^- \to e^+e^-\mathcal{P}$ under the event selection
similar to that of the single-tag experiments.
The pattern of data/simulation agreement is the 
same as for the form factors and this is 
an important consistency check.

Recently, {\tt EKHARA} was used by KLOE-2 Collaboration~\cite{Babusci:2011bg}
for the feasibility studies of $e^+e^- \to e^+e^-$ $\pi^0$.
On the basis of simulation,
a per cent level of precision is foreseen for the
measurement of $\Gamma_{\pi^0 \to \gamma\gamma}$ by the
KLOE-2 experiment at Frascati.
KLOE-2 experiment can also perform the first measurement  
of pion transition form factor $F_{\pi^0\gamma^*\gamma}(Q^2)$ 
in the space-like region in the vicinity of the origin: 
$0.01 < Q^2 < 0.1~\mbox{GeV}^2$.
The impact of the proposed measurements 
on the evaluation of the Standard Model prediction 
for the anomalous magnetic moment of the muon, $a_\mu$, was estimated~\cite{Babusci:2011bg}.


\section{Summary}
\label{sec:summary}
 A brief review of the Monte Carlo event generator {\tt EKHARA} is presented
 with the focus on the physics motivation and several new features.
 It can be used for the simulation of the processes
 $e^+e^- \to e^+e^-\mathcal{P}$ and $e^+e^- \to e^+e^- \pi^+\pi^-$.
 The channels $e^+e^- \to e^+e^- \pi^0$, $\eta$, $\eta^\prime$ 
 are important for $\gamma^{*}\gamma^{*}$ physics programs of the
 modern experiments (e.g., KLOE-2 and BES~III) 
 and can be used for the simulation of the 
 two-photon form factor measurements at meson factories.
 The program is supplied with the form factors for  
 $\pi^0$, $\eta$ and $\eta^\prime$ mesons in agreement with the modern
 data on the transition form factors and corresponding differential
 cross sections $\frac{d\sigma}{d Q^2}$.

 The future plans of {\tt EKHARA} development include the 
 electromagnetic radiative corrections
 for the channel $e^+e^- \to e^+e^-\mathcal{P}$.
 After the inclusion of the radiative corrections we plan
 to perform a detailed comparison of our generator with 
 {\tt GGRESRC}~\cite{Druzhinin:2010er}, 
 which was used by the BaBar experiment~\cite{Aubert:2009mc,:2011hk}.
 The major update of the $e^+e^- \to e^+e^- \pi^+\pi^-$ channel
 in context of possible $\gamma^{*}\gamma^{*}$ physics motivation
 is also foreseen.

\section*{Acknowledgments}

This research was partly supported by 
Polish Ministry of Science and High Education
from budget for science for years 2010-2013: grant number N~N202~102638,
by Sonderforschungsbereich SFB1044 of the Deutsche Forschungsgemeinschaft
and by
National Academy of Science
of Ukraine contract $50/53$---$2011$.
This work is a part of the activity of the ``Working Group on Radiative
Corrections and Monte
Carlo Generators for Low Energies''~\footnote{\href{http://www.lnf.infn.it/wg/sighad/}{http://www.lnf.infn.it/wg/sighad/}}.








\end{document}